# Employing p-CSMA on a LoRa Network Simulator

Nikos Kouvelas, Vijay Rao, R.R. Venkatesha Prasad
Embedded Software, EEMCS, TU Delft, the Netherlands

## ABSTRACT

Low-Power Wide-Area Networks (LPWANs) emerged to cover the needs of Internet of Things (IoT)-devices for operational longevity and long operating range. Among LPWANs, Long Range (LoRa) WAN has been the most promising; an upcoming IoT protocol, already adopted by big mobile operators like KPN and TTN. With LoRaWANs, IoT-devices transmit data to their corresponding gateways over many kilometers in a single hop and with 1% duty-cycle. However, in a LoRa network, any device claims the channel for data-transmission without performing channel-sensing or synchronization with other devices. This increases humongously the number of collisions of information-packets when the number of IoT-devices that are connected per gateway increases.

To improve the utilization of the channel, we propose the application of persistent-Carrier Sense Multiple Access (p-CSMA) protocols on the MAC layer of LoRaWANs. In this manuscript, we report on the initial design of a p-CSMA component for the simulation of LoRa networks in $ns3$[1]. In particular, the classes adding p-CSMA functionality to the IoT-devices are presented. Additionally, the dependencies and relations between these classes and an existing LoRaWAN module on which they apply are detailed. Further, we evaluate this new p-CSMA LoRaWAN module in terms of Packet Reception Ratio (PRR) by simulating LoRa networks. The current report is the first step in the creation of a holistic p-CSMA module, directed to support network-researchers and connoisseurs in simulating all aspects of LoRa networks in $ns3$.

## KEYWORDS

ns3, LoRaWAN, p-CSMA, channel sensing, persistence, hidden terminals, scalability

## 1 INTRODUCTION

The current trend of Cyber-Physical Systems, i.e., distributed systems of autonomous sensors and actuators operating in independent localities, is growing with a tremendous rate. The Internet of Things (IoT) paradigm and the advent of the Tactile Internet bridge the gap between the physical world of sensors and actuators and the cyber world which defines the communication between them, by providing ubiquitous connectivity ('smart' things) and minimal communication delay (round-trip time between sensors and actuators). This is bringing about the next Industrial Revolution, called Industry 4.0, wherein most operations will be automated with sensors and actuators deployed in orders of hundreds of thousands. Currently, most of these devices base their operational longevity on batteries and they consume energy, especially to sense and communicate wirelessly. Therefore, eventually they die out and need to be replaced, provoking environmental and economic damage.

To improve their sustainability, many distributed systems of IoT-devices communicate using Long Range Wide Area Networks (LoRaWANs); a technology intended for low bit-rate, wireless communications among distant, resource-constrained devices. Operating in a star-cellular topology, LoRa end-devices transmit their information directly to their gateways, which are connected to the Internet, with 1% duty cycle. In the PHY layer, LoRa-packet payloads are derived as series of chirps, created by a linear frequency modulation technique, named Chirp Spread Spectrum (CSS). LoRa offers six different Spreading Factors (SFs) for the aforementioned CSS (SF7 to SF12), modulating signal-symbols into $2^{SF}$ chips. LoRa SFs manifest orthogonality, allowing messages modulated by different SFs to be received on the same channel simultaneously (eight channels per gateway). Since this work is focused on simulating the MAC layer of LoRaWANs, if the reader needs thorough information on LoRa(WAN), (s)he is redirected to [1].

### 1.1 The scalability constraints of the current MAC layer of LoRaWAN

Since Industry 4.0 brings a multitude of 'smart' sensors and actuators, LoRa networks need to be scalable; able to employ dependably hundreds of IoT-devices per gateway. This is the only economical way to sustain the systems of sensors and actuators serving the 'smart-x' applications of the future. However, the scalability potential of LoRaWANs is hindered by their current MAC layer protocol, which operates under the ALOHA-principle; when a device has a packet to transmit, it transmits directly. Current works show that the relation between throughput and traffic for devices of the same SF follows the ALOHA-curve, with $\sim 0.184$ packets received per packet-time when the channel is saturated (i.e., 0.5) [2]. The eight-point orthogonality that is offered by LoRa-receivers–although increasing the throughput when devices of different SFs are used– obviously cannot mitigate the LoRaWAN channels saturating in low throughput values. Thus the employment of more gateways is mandated.

### 1.2 Contributions

To this point, we propose the application of persistent-Carrier Sense Multiple Access (p-CSMA) protocols on the MAC layer of LoRa networks; first, to deal with the lack of channel sensing, which contributes the most regarding packet collisions, and second, to maximize the channel utilization. In particular, each IoT-device senses the wireless channel before transmitting, avoiding simultaneous transmissions with other devices. Once the channel is sensed idle, the device is free to transmit. However, when an IoT-device performs channel sensing, the procedure can reveal solely devices located in the sensing-vicinity of the sensing-device. Therefore, the packet collisions that are provoked by devices located out of the aforementioned vicinity (i.e., coverage range) cannot be avoided. This condition, known as 'hidden-terminal', imposes limitations on any network operating under channel-sensing protocols. In p-CSMA, the value of persistence, $p$, dictates the probability by which a device will transmit once it senses the channel being idle [3]. This persistence value can be adjusted to avoid the simultaneous transmission from IoT-devices 'hidden' to each other.

---
[1]https://github.com/nikoskouvelas/lorawanpCSMA

In this manuscript, we detail the classes that are created in order to simulate p-CSMA functionality in LoRaWAN modules in *ns3*, and we define their interactions and dependencies. Network Simulator 3 (ns3) is an open-source simulator of considerable utilization in networking research. The p-CSMA functionality is approached solely in terms of uplink messages, i.e., messages transmitted from the IoT-devices and received by the gateways. Specifically, our contributions are the following:

- we developed classes simulating channel-sensing, hidden-terminals, persistent channel-claiming,
- we connected these classes with an already operating module [2] to create a new `lorawan` module, and
- we measured the Packet Reception Ratio (PRR) of this new module by simulating scenarios with different groups of IoT-devices and one gateway.

The rest of this work is organized as follows: Section 2 refers to related works. Section 3 discusses thoroughly the design and structure of the p-CSMA component. In Section 4, the results of the simulation are presented. In Section 5, the future extensions of our current model are stated. Finally, Section 6 concludes this work.

## 2 LoRaWAN MODULES IN NS3

*ns3* simulates the operation of network-topologies as a discrete series of events taking place in a forward time-manner. Being open-source, object-oriented (written in C++), and by offering attributes like configurable tracing system and automatic memory management, *ns3* provides many knobs for researchers to leverage the specifics of any networking scenario in order to derive meaningful results [4].

There are two *ns3* modules modeling adequately a LoRa network of IoT-devices and gateways under a simple LoRaWAN network server [2, 5]. Both these `lorawan` modules start from modeling the specifics of transmission/reception on the PHY layer and continue with the manner in which information payloads are communicated among devices and gateways (MAC layer) to be finally pushed to the Internet (application layer).

In [5], Van den Abeele et al. evaluated the impact of important network parameters on the scalability of LoRaWAN, by using the *ns3* `lorawan` module that they created. In particular, the authors focused on increasing the network traffic (through the number of devices), changing the number and position of network gateways, using gateway-acknowledgements upon transmissions, and assigning different data-rates on the end-devices (through changing their SFs).

In [2], using their *ns3* `lorawan` module, Magrin et al. evaluated the performance of LoRaWAN, seen as the enabler of the IoT-paradigm in the dense IoT-deployment of a smart city scenario. To this point, the authors related network throughput and reliability of coverage to the number of devices/gateways, duty cycle limitations, and variety of SFs among the IoT-devices.

In terms of the MAC layer, regardless the simulation specifics in each of the aforementioned related works, the medium access is assumed happening by following the ALOHA protocol. In this manuscript, based on the open source code of [2], we added several new classes and manipulated specific existing ones in order to employ the p-CSMA protocol. Additionally, based on the SNR limits that dictate the changing of the SF among devices in [2], we modeled the conditions under which two terminals are 'hidden' of each other.

## 3 p-CSMA FRAMEWORK FOR LoRaWAN MODULES

The high-level structure, the design goals, and the assumptions taken to create the p-CSMA framework are presented in this section.

### 3.1 Design Goals

The scope of this framework is to provide a complete model of the p-CSMA protocol, tailored on the specifics of a LoRa network. Device-wise, channel claiming should involve the following functionalities:

- sensing the channel to find it idle/occupied from non-hidden devices,
- if idle, capturing the channel and start transmitting,
- if occupied backing-off and continue sensing the channel, and
- retry capturing the channel with a certain persistence when it is sensed idle.

This work, being the first step towards a holistic p-CSMA framework, realizes p-CSMA under an already given persistence $p$-value (by the user). In the future, we intend to develop algorithms on the automatic update of $p$-values. Further, a model for energy consumption or generation (by harvesting) throughout the simulation is intended. The future coding attempts will also focus on reducing the overall code size and the simulation overhead.

### 3.2 Assumptions

The classes comprising our proposed *ns3* framework are based on the assumptions that follow:

- Two channel-states are assumed, idle (0) and occupied (1), following the scheme: idle → occupied → idle.

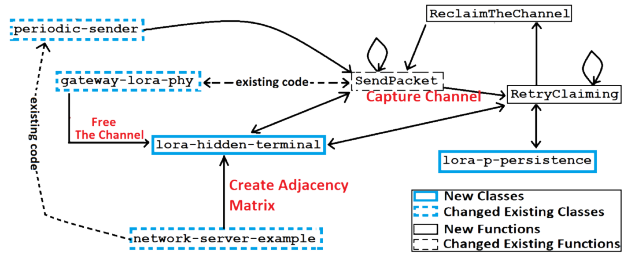

**Figure 1: High-level structure of the *ns3* framework.**

- All devices, regardless their distance from the gateway are informed instantaneously on the (change of) state of the channel (no round-trip delays).
- If multiple devices attempt to transmit simultaneously, the channel, if idle, is claimed in a FIFO manner.
- In theory [3], when the channel is sensed as occupied, the device continues sensing it until it becomes idle. In this work, we assume sensing intervals that last for half of the time needed by a newly transmitted payload to reach its gateway and be demodulated[2]. Obviously, this deterministic value can be changed to encapsulate more (or less) energy efficient strategies in future energy classes, i.e., increasing the frequency of sensing increases the energy spent device-wise.
- The existing `lorawan` module [2] models broadcast transmissions from the end-devices. Therefore, when a device transmits, along with the gateway, its packet is also delivered to all the devices in its coverage range. However, each end-device is set to drop packets originating from other end-devices.
- The current *ns3* framework is focused solely on stationary devices. In future scenarios, mobile devices will be also involved in order to explore the changes that their mobility brings in the specifics of transmission.

### 3.3 General Structure

Fig. 1 presents a rough overview of the proposed *ns3* framework on LoRaWAN. The newly added classes and critical interconnecting functions are presented along with the existing ones that were changed by the authors. For thorough information on the whole `lorawan` model (before the contributions of this work), the reader is redirected to the GitHub code of [2].

#### 3.3.1 Network Server Example.
The class `network-server-example` is the control panel of the simulation. By this class, the user can decide the channel propagation model, fading characteristics, and total number of gateways/end-devices, along with their PHY and MAC characteristics. Also, the period of transmission of each device and the total running time of the simulation are decided here.

#### 3.3.2 Periodic Sender.
The existing class `periodic-sender` manages the periodic transmissions of the information-packets by the end-devices, according to the periodicity-data that is received by the `network-server-example`. In particular, each device calls the class `lora-hidden-terminal` to reveal which other devices are positioned in its transmission range, and sense if the channel is occupied by one of them. If the channel is idle, `SendPacket` captures it to transmit the information-packet of the end-device, rescheduling at the same time the next transmission of the packet by the same end-device. If the channel is occupied, `SendPacket` calls the function `RetryClaiming`, which continues sensing the channel frequently (i.e., half the time needed by a newly transmitted payload to reach its gateway and be demodulated.). Once the channel becomes free, `RetryClaiming` calls the class `lora-p-persistence` to define if transmission is allowed according to the imposed *p*-value. In case *p*-value permits, `RetryClaiming` calls the function `ReclaimTheChannel`, which triggers `SendPacket` again in order for the device to send the corresponding payload.

#### 3.3.3 Lora Hidden Terminal.
The class `lora-hidden-terminal` initializes and updates the condition of the channel. In addition, it replies regarding the current condition of the channel, i.e., if the channel is occupied by any device in the vicinity of the sensing-device or if the channel is idle. Therefore, this class grants or declines the permission of capturing the channel to an end-device. Due to the above, `lora-hidden-terminal` interacts with the function `SendPacket` to inform on the channel condition and it is called by the class `gateway-lora-phy` whenever the channel must become idle. Furthermore, before the simulation starts, `lora-hidden-terminal` is called by `network-server-example` to create the *adjacency matrix*, dictating which devices are in the coverage range (i.e., transmittance vicinity) of each end-device.

#### 3.3.4 Lora p-persistence.
The class `lora-p-persistence` is called from the function `RetryClaiming` to check a device's *p*-value of persistence, in order to grant (or not) the permission to transmit in the channel. `lora-p-persistence` includes a group of functions which update the *p*-value for a device. Then, using a `rand` function, it is dictated if the device will transmit.

#### 3.3.5 Gateway lora phy.
The class `gateway-lora-phy` is changed by this work, to update the state of a channel from occupied (1) to idle (0) in any of the following cases:

- the packet was dropped because its SNR is lower than the receiving threshold of the gateway (i.e., sensitivity),
- the packet was dropped because all the eight orthogonal receiving paths in the channel frequency were captured,
- the packet collided with another packet, originating from a node hidden to the one that transmits, and
- the packet was received correctly.

---

[2] $19 byte$ payloads are used, with $\sim (0.51/2)s$ being the sensing interval

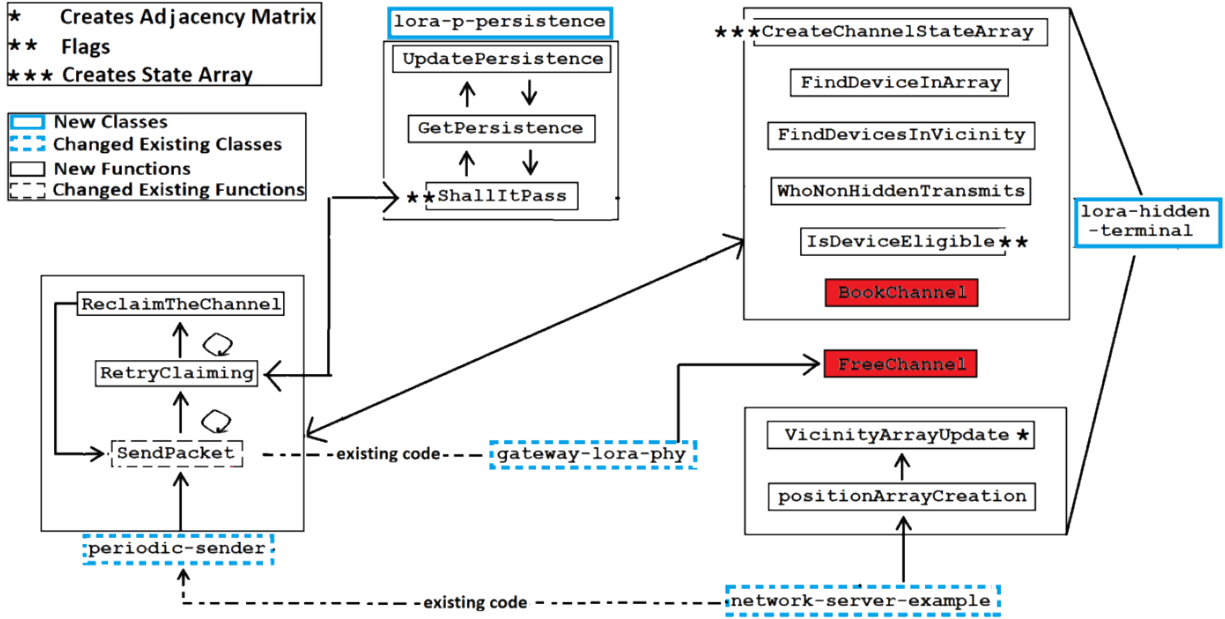

Figure 2: Class Diagram of the *ns3* p-CSMA framework.

## 3.4 Framework Specifics

At this point, we analyze thoroughly the two new classes created for the p-CSMA functionality, `lora-hidden-terminal` and `lora-p-persistence`. Their scope and dependencies with other parts of the `lorawan` module are specified. A detailed, structural diagram of the *ns3* p-CSMA framework, involving the aforementioned classes and their functions, is presented in Fig. 2.

*3.4.1 lora-hidden-terminal.* This class includes three different blocks of functions, interacting with three different classes of the `lorawan` module, as seen in Fig. 2. In the initialization phase, before the actual start of the simulation, `network-server-example` calls the function `positionArrayCreation`, which takes as input the 3D-vectors denoting the position of every end-device and forwards them to the function `VicinityArrayUpdate`, which in turn creates the *adjacency matrix*, simulating the 'hidden-terminal' condition (see the '*' in Fig. 2). In particular, since every end-device is stationary and has a specific SF, the maximum distance that can be reached by its transmitted packets is known from the beginning of the simulation. By comparing the Euclidean distance of every pair of devices to the already known maximum distances, it is revealed which devices are in the coverage range of every specific end-device. During the simulation, two functions of the class `lora-hidden-terminal` change the state of the channel; the `BookChannel` and the `FreeChannel`[3]. But the state of the channel has to be initialized first, and then needs to be updated whenever the channel gets captured or becomes idle. At the beginning of the simulation, the function `CreateChannelStateArray` is triggered to initialize an array that holds 0 for every end-device (i.e., the number or devices denotes the length of the array). This means that the channel begins as idle regarding all the declared end-devices (see the '***' in Fig. 2). From then onwards, when there is a packet from an end-device to be transmitted, the functions `FindDeviceInArray` and `FindDeviceInVicinity` are called by the `periodic-sender` to check which specific devices are in the transmitting vicinity of the end-device that requires access to the channel. Next, the function `WhoNonHiddenTransmits` checks the *channel state array* to reveal if among the aforementioned devices-in-vicinity there is at least one that transmits. According to the output, the function `IsDeviceEligible` marks a flag-variable (see the '**' in Fig. 2). If this flag is 0, then the function `BookChannel` is triggered, otherwise the function `RetryClaiming`, that was previously discussed, is called. In case the `BookChannel` is called, it updates the value of the corresponding position of the *channel state array*, idle (0) → occupied (1), and the `SendPacket` function continues further with the transmission of the packet. For an occupied state of the *channel state array* to become idle again, the function `FreeChannel` needs to be triggered from the class `gateway-lora-phy`. This can happen if the packet is dropped, collided, or accepted, as explained in the previous section.

*3.4.2 lora-p-persistence.* This class includes three different functions triggered in a serial manner to grant (or refuse)

---
[3]In the code they are named `BookChannelStateArrayCondition` and `FreeChannelStateArrayCondition`

the allowance to an end-device to use the channel for transmission, as seen in Fig. 2. The class `lora-p-persistence` is called specifically by the function `RetryClaiming`, because the concept of persistence in transmissions is introduced in the p-CSMA only when a device backed-off from transmitting in the channel and tries to reclaim it. At that point, the function `ShallItPass` is called to compare the result of a `rand` with the $p$-persistence value of the device, and mark a flag accordingly (0 or 1) to permit or not the transmission (see the '**' in Fig. 2). The $p$-value is obtained by calling the function `GetPersistence` which keeps an array with every $p$-value per end-device in the system. The $p$-values in the aforementioned array are updated by calling the function `UpdatePersistence`. The method in which $p$-values are updated is not finalized yet, thus at the moment, the user can choose the $p$-value (s)he wishes for each of the devices in the system. For example, the $p$-values can be updated by the user at will by observing the results regarding the Packet Reception Ratio (PRR).

## 3.5 Discussion

Fig. 2, by presenting analytically the *ns3* framework, encloses the three specific conceptual parts of a p-CSMA. Specifically:

- The Multiple-Access (MA) part is realized by the class `periodic-sender`. In [2], this class was involving only the function `SendPacket`, which simply occupied the channel whenever an end-device had a payload to transmit. In this work, `periodic-sender` is updated to interoperate with the other classes of MAC to allocate the channel according to the results of channel-sensing and persistence.
- The Channel-Sensing (CS) is realized by the class `lora-hidden-terminal`, which creates the *adjacency matrix* of (non)-hidden devices and updates the state of the channel that every end-device senses (i.e., idle or occupied).
- The effect of the $p$-persistence on the claiming of the channel by the end-devices is realized by the class `lora-p-persistence`, which updates the $p$-value and grants or declines transmission to any end-device that sensed the channel as idle.

## 4 SIMULATION

In this section, the proposed p-CSMA framework is used to investigate Packet Reception Ratio (PRR) in LoRa networks.

### 4.1 Setup

The following parameters were tuned:

- number of devices,
- transmission period ($100s$, $200s$, $300s$, $400s$, and $500s$),
- persistence values,
- areas with terminals that are hidden from each other, and
- Spreading Factors of devices (only SF8 or SF8, SF9, and SF10).

One hour of operation is simulated. One gateway is assumed. The devices are considered stationary and equally distributed regarding transmission periods, number of devices in each area that is 'hidden' from the others, and SFs. Eight orthogonal receiving windows are used on a sole receiving frequency of $868.1 MHz$.

### 4.2 Results

At first, even without presenting the case of a classic ALOHA-LoRaWAN, it is obvious that the PRR is improved by using this p-CSMA *ns3* framework. Although the transmission periods are chosen in a way to provoke numerous collisions during the simulation, many of these collisions are evaded directly by using channel-sensing. Further, as seen in Fig. 3 and Fig. 4, when lower $p$-values are used, more packets are received correctly. The low probability with which each device claims the channel prevents many collisions by excluding devices from transmitting. In addition, because of the orthogonality of the eight receiving paths, it is observed that by using three different SFs (8, 9, and 10) instead of one (only 8), relatively higher PRR values are reached (cf. Fig. 4), approaching even the value of PRR= 1 for low number of devices. As expected, increasing the number of devices produces more traffic in the network, leading to lower PRR values. However, in certain cases of $p = 0.25$, the PRR for 80 devices is higher than the one for 60 devices. This implies underutilization of the channel for this specific persistence value. Regarding the simulation of groups of hidden terminals in two or three different areas of the topology, a clear outcome cannot be produced by Fig. 3 and Fig. 4. In future versions of this work, many more devices will be simulated in order to see how the persistence values affect the packet reception in large-scale networks with groups of devices located out of the coverage range of others.

## 5 FUTURE EXTENSIONS OF THIS FRAMEWORK

### 5.1 A class that changes persistence dynamically

In this work, the $p$-value is chosen by the user. However, a short-term extension of this framework will include the design of a sophisticated `UpdatePersistence` function inside the `lora-p-persistence` class in order to achieve maximized utilization of the channel by the end-devices or to prioritize specific (groups of) end-devices.

### 5.2 A battery module

LoRaWANs are created to offer long distance communication among energy constrained, battery powered IoT-devices. Thus, in order to complete the picture of a LoRa network, the lifetime of the participating devices and gateways should be simulated. Therefore, a framework simulating the energy that is spent in transmitting, sensing, and receiving will be created to operate in parallel with the current p-CSMA framework. Additionally, since energy harvesting is currently utilized in many ecosystems of IoT-devices, our future battery module will also cover energy scavenging from renewable energy sources.

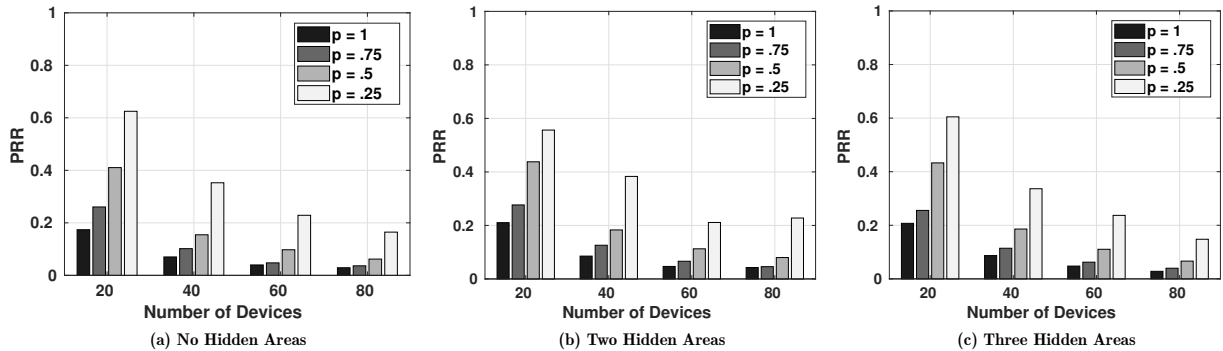

Figure 3: Packet Reception Ratio, Spreading Factor of 8.

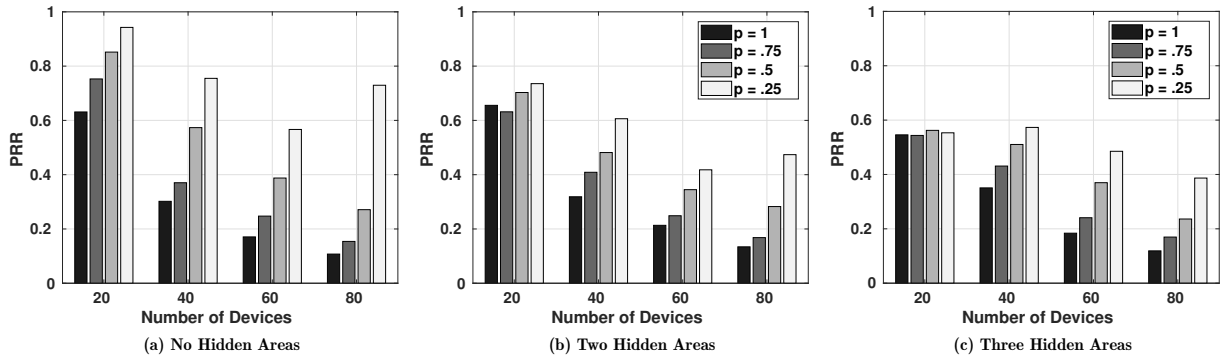

Figure 4: Packet Reception Ratio, Spreading Factors of 8, 9, and 10.

## 6 CONCLUSION

In this manuscript, we presented a modular and extensible framework that simulates p-CSMA on the MAC layer of LoRaWAN in *ns3*. The designed framework can be used to simulate channel sensing, hidden devices and transmission back-offs under different persistence values. A high-level structural view of our work was presented, followed by an explanatory class diagram. Further, through simple simulation examples, the Packet Reception Ratio of small LoRa Networks was examined under different number of (hidden) devices, persistence values, and Spreading Factors. We consider the application of p-CSMA on the MAC layer of LoRaWAN as an important step towards scalability for this type of low-power networks. Thus, this framework and its future extensions are aimed to be valuable tools in the hands of network engineers and researchers in the field of telecommunications.

## ACKNOWLEDGEMENT

SCOTT http://www.scott-project.eu has received funding from the Electronic Component Systems for European Leadership Joint Undertaking, grant agreement No 737422. This joint undertaking is supported from the EU Horizon 2020 research and innovation program and Austria, Spain, Finland, Ireland, Sweden, Germany, Poland, Portugal, Netherlands, Belgium, Norway.